\documentclass[aps,prl,twocolumn,amsmath,amssymb,superscriptaddress,floatfix,longbibliography]{revtex4-2}
\pdfoutput=1
\usepackage{mathrsfs}
\usepackage{amsmath}
\usepackage{mathtools}
\usepackage{braket}
\usepackage{amsfonts}
\usepackage{float}
\usepackage{graphicx} 
\usepackage[caption=false]{subfig}
\usepackage{bm} 
\usepackage{xcolor}
\usepackage{epstopdf}
\usepackage{comment}
\usepackage{soul}
\usepackage{multirow}
\usepackage{physics}
\usepackage[normalem]{ulem}
\usepackage{nicefrac, xfrac}
\usepackage{hhline}
\usepackage{simpler-wick}
\usepackage{titlesec}
\usepackage{cancel}

\usepackage{xr-hyper}
\usepackage{hyperref} 

\titlespacing*{\paragraph}
{0pt}{3.25ex plus 1ex minus .2ex}{1.5ex plus .2ex}

\newcommand\ok[1]{\textcolor{black}{#1}}

\newcommand{\Sum} [2] {\the\numexpr #1 + #2 \relax \\}

\begin{document}
\title{Beyond Topological Invariants: Order Parameters from Dominant Fock-state Patterns}


\author{ Tsz Hin Hui }
\affiliation{ \textit Department of Physics, City University of Hong Kong, Kowloon, Hong Kong, China}

\author{ Xiaodan Xia }
\affiliation{ \textit Department of Physics, City University of Hong Kong, Kowloon, Hong Kong, China}

\author{Pedro D. Sacramento}
\affiliation{ \textit CeFEMA, Instituto Superior T\'ecnico, Universidade de Lisboa, Av. Rovisco Pais, 1049-001 Lisboa, Portugal}

\author{ Wing Chi Yu }
\email{wingcyu@cityu.edu.hk}
\affiliation{ \textit Department of Physics, City University of Hong Kong, Kowloon, Hong Kong, China}

\date{\today}

\begin{abstract}
We introduce a general scheme for constructing order parameters (OPs) by extracting generic patterns from the dominant Fock states of many-body ground states. While topological phases are traditionally characterized by non-local invariants, we demonstrate that our real-space OPs provide a more refined classification. In the extended Su-Schrieffer-Heeger model, we show that the standard winding number is insufficient to fully distinguish all phases; our OPs reveal a hidden sub-structure where each topological sector splits into two distinct phases. Beyond identifying the phase boundaries, these OPs quantify the depth of a phase, and remain robust in characterizing transitions in disordered systems. Furthermore, our approach provides a practical finite-size diagnostic for the Berezinskii–Kosterlitz–Thouless  transition in the interacting spin-1/2 XXZ model. The presented framework offers a broadly applicable tool for uncovering the phase diagrams of diverse interacting and non-interacting quantum many-body systems.
\end{abstract}

\maketitle


\emph{Introduction.---} The construction of an appropriate order parameter (OP) is a cornerstone of many-body physics, yet there is no universal recipe; proposed forms often rely on physical intuition. Over the past decade, two main methods have been developed: (i) a variational approach that detects the ``quasi-degenerate" states indicative of the spontaneous symmetry breaking \cite{furukawa2006,cheong2009,
henley2014}, and (ii) approaches that utilize the spectrum of the reduced density matrix to construct OPs \cite{gu2013,yu2016,yu2019,arraut2024,yu_gu2016}. The second approach, despite being non-variational, the analysis depends on the choice of a suitable subsystem, which may not be straightforward to identify in systems with long-range couplings. 

In this letter, we propose a general scheme for constructing the OP. The target is to find an order parameter $\hat{O}$, such that its ground state (GS) expectation value $\langle\hat{O}\rangle\equiv\langle\text{GS}|\hat{O}|\text{GS}\rangle$ gives non-zero value in its own phase, and returns zero (or at least close to zero \cite{comment1}) otherwise. The fundamental idea of our scheme is to extract the generic pattern in the dominant Fock states of the GS to construct the OP. Mathematically, the GS reads
\begin{align}
    \ket{\text{GS}} = \sum_{n_1{}_, n_2{}_,\cdots{}_, n_{L}}\xi_{n_1 n_2\cdots n_{L}}\ket{n_1 n_2\cdots n_{L}},
    \label{eqn:fock_scheme}
\end{align}
where $\ket{n_1 n_2\cdots n_{L}}$ is the Fock state, with $n_j$ being the occupancy on site $j$. In fermionic models, $n_j$ can be 0 or 1. In spin-$s$ model, this can be any value in $-s,-s+1,\cdots,s-1,s$. We analyze the Fock states that have dominant weights to infer the form of $\hat{O}$ that gives $\hat{O}\ket{n_1 n_2\cdots n_{L}}\neq 0$ on the dominant Fock states $\ket{n_1 n_2\cdots n_{L}}$. 
The number of dominant Fock states in consideration need not to be only one but can have multiple of them depending on the situation.

The reason for focusing on the dominant Fock states is two-fold. First, it is easier to handle. The total number of Fock states scales as $l^L$, where $l$ is the dimension of the local Hilbert space and $L$ is the total number of sites, which is too large to analyze. Second, it is expected that the Fock states with large weights are more representative of the state (or the phase). From the standpoint of quantum theory, Eq. (\ref{eqn:fock_scheme}) means that the state $\ket{n_1 n_2\cdots n_{L}}$ has a probability of $|\xi_{n_1 n_2\cdots n_{L}}|^2$ being observed when the $ \ket{\text{GS}}$ collapse under a measurement. Therefore, large-weight Fock states are more probable to be observed. Some remarks on the scheme are given in the Supplemental Material (SM) \cite{scheme_sm}. 

Below, we illustrate the scheme to derive the OPs in the extended Su-Schrieffer-Heeger (ESSH) model. These OPs not only recover known phase boundaries but also refine the phases classification beyond the winding number, and they remain effective for detecting transitions in disordered settings. Furthermore, we show that the scheme provides a robust diagnostic for the Berezinskii–Kosterlitz–Thoules (BKT) transition in the interacting spin-$\frac{1}{2}$ XXZ model.

\emph{ESSH model.---}The SSH model is one of the prototypical models of topological insulators. In recent years, there has been growing interest in its extension that includes hopping beyond the nearest-neighbors \cite{maffei2018,wong2024,Rufo2019,Li2018,Li2014,Chang2025,Ghosh2023,Perez-Gonzalez2019,Cinnirella2024}. One reason is the increasing number of experimental platforms reporting rich topological phases induced by long-range hoppings \cite{Li2024,Liu2023,Wang2023,Leefmans2022,he2026}, 
opening avenues to engineer novel artificial quantum phases of matter. 

\begin{figure}
    \centering
    \includegraphics[width=8cm]{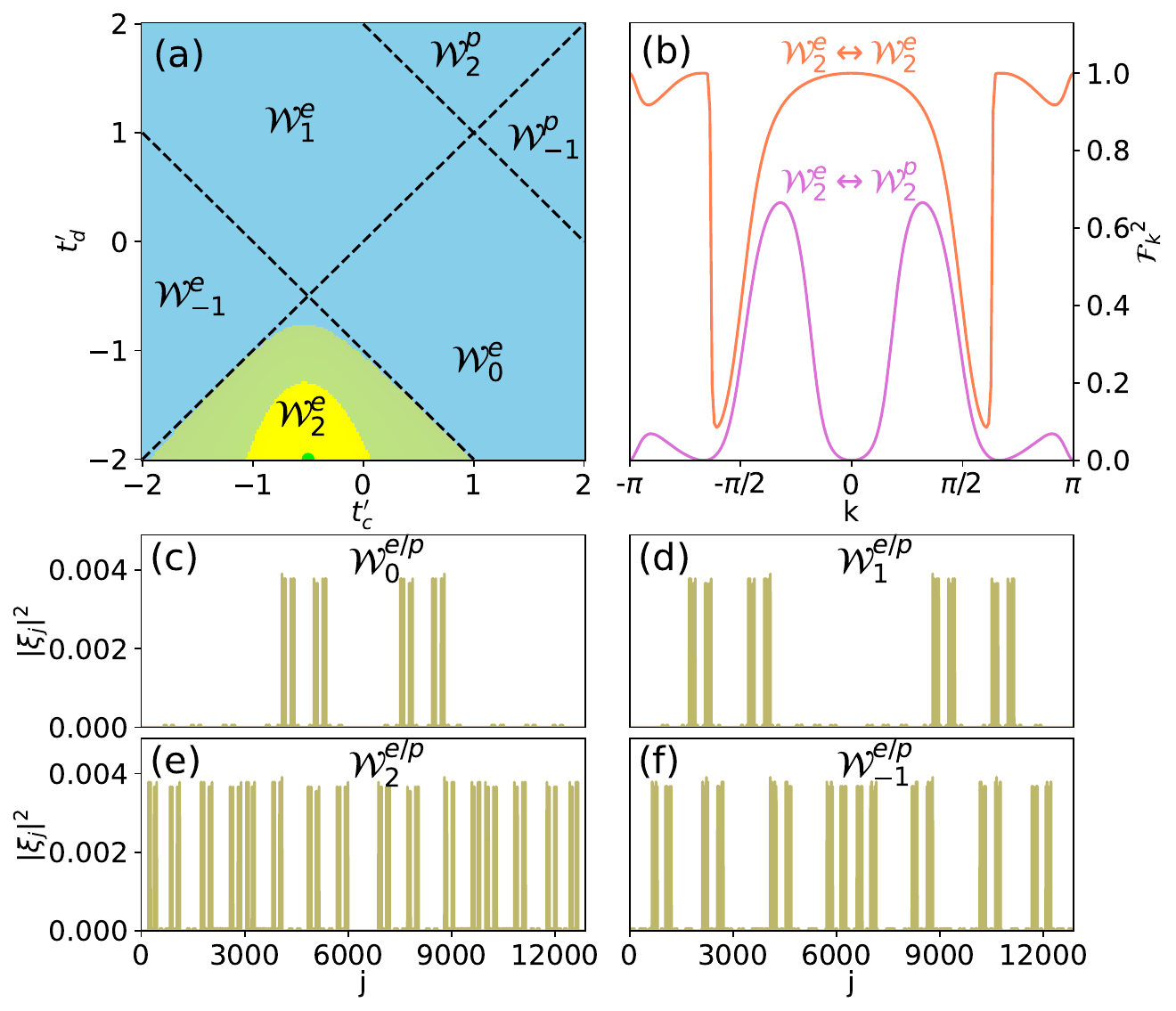}
    \caption{(a-b) shows the fidelity $\mathcal{F}(\vec{\lambda},\vec{\lambda}')$. 
    (a) The green dot indicates the reference point where $\vec{\lambda}=(t_a,t_b,t_c,t_d)=(-1,-1,-0.5,-1.8)$ and $\vec{\lambda}'=(t_a',t_b',t_c',t_d')=(-1,-1,t_c',t_d')$. The dashed lines denote the phase boundaries determined by the winding numbers, and the phases are labeled as $W_m^i$ with $m$ being the winding number and $i$ being either ``e" or ``p". Blue (yellow) indicates $\mathcal{F}=0$ ($\mathcal{F}=1$). The larger (smaller) yellow region with dimmer (sharper) color corresponds to the $\mathcal{F}=1$ for system size of $N=5000$ (50000) \cite{Andersonnote}; 
    (b) $\mathcal{F}_k^2(\vec{\lambda},\vec{\lambda}')$ at the thermodynamic limit for two specific cases, where $\vec{\lambda}=(-1,-1,-0.5,-1.8)$ for both cases. The coral [pink] curve correpsonds to $\vec{\lambda}'=(-1,-1,-0.5,-0.6)$  [$\vec{\lambda}'=(-1,-1,1,2)$]. 
    (c-f) show the squared modulus of the weight  $|\xi_j|^2$ (see Eq. (\ref{eqn:fock_scheme})) of the Fock states for the GSs from deep inside different phases: (c) $\mathcal{W}_0^{e}$ phase where $(t_a,t_b,t_c,t_d)=(-1.6,-0.4,0,0)$; (d) $\mathcal{W}_1^{e}$ phase where $(t_a,t_b,t_c,t_d)=(-0.4,-1.6,0,0)$; (e) $\mathcal{W}_2^{e}$ phase where $(t_a,t_b,t_c,t_d)=(0,-0.5,0,-2)$; (f) $\mathcal{W}_{-1}^{e}$ phase where $(t_a,t_b,t_c,t_d)=(0,-0.5,-2,0)$. Note that (c) and (d) are the conventional SSH model. Here, $N=8$. }
    \label{fig:fidelitymap}
\end{figure}

The Hamiltonian of the ESSH model reads \cite{maffei2018,wong2024}
\begin{align}
    H_\text{ESSH} &= \sum_{j = 1}^{N} ( t_ac_{j,A}^\dagger c_{j,B} + t_bc_{j,B}^\dagger c_{j + 1,A} \nonumber \\
    & \qquad + t_cc_{j,A}^\dagger c_{j + 1,B} + t_dc_{j,B}^\dagger c_{j + 2,A} + \text{h.c.} ).
    \label{eqn:H_SSH}
\end{align}
We consider half-filling and periodic boundary condition (PBC). In the original SSH model ($t_c=t_d=0$), $|t_a|>|t_b|$ ($|t_a|<|t_b|$) corresponds to topologically trivial (non-trivial) phase, characterized by $\mathcal{W}_0$ ($\mathcal{W}_1$). Here, $\mathcal{W}_m$ denotes the phase with winding number $\mathcal{W}=m$, for $m\in\mathbb{Z}$. When $t_c,t_d\neq 0$, $\mathcal{W}_{-1}$ and $\mathcal{W}_{2}$ phases can emerge, which are also topological phases. As argued in Ref. \cite{wong2024}, when $|t_c|$ ($|t_d|$) is larger than the remaining model parameters, the system is in the $\mathcal{W}_{-1}$ ($\mathcal{W}_2$) phase. This is evident from Fig. \ref{fig:fidelitymap}(a), where the phase diagram of $t_a=t_b=-1$ as a function of $t_c$ and $t_d$ is shown. The superscript ``e'' and ``p'' is explained later. 

\emph{Insufficiency of the winding number.---}It is a general belief that the winding number captures all the phases in the ESSH model. Here, we show that distinct phases can share the same winding number, and we introduce a superscript to distinguish these phases for each winding number sector, namely, $\mathcal{W}_m^e$ and $\mathcal{W}_m^p$. Parts of the literature have implicitly treated them as identical ones \cite{song2014,lin2021,Sadrzadeh2021,Liu2017,Li2014,Rufo2019,wong2024,maffei2018}. 

Figure \ref{fig:fidelitymap}(a) shows the GS fidelity $\mathcal{F}(\vec{\lambda},\vec{\lambda}')=\bigl|\langle\Psi_0(\vec{\lambda})|\Psi_0(\vec{\lambda}')\rangle  \bigl|$ with $\lambda$ chosen in the $\mathcal{W}_2^e$ phase (see the green dot) and $\lambda'$ is varied over the phase diagram. Physically, it measures the GS similarity between the $\mathcal{W}_2^e$ phase and all the points in the phase diagram. We find that the $\mathcal{W}_2^p$ phase always exhibit zero fidelity with $\mathcal{W}_2^e$ phase. Such zero in the fidelity can be explained by Fig. \ref{fig:fidelitymap}(b) (see pink curve). It displays $\mathcal{F}_{k}=\bigl|\bigl\langle 0 \bigl|\beta_k^{}(\vec{\lambda})\beta_k^\dag(\vec{\lambda}') \bigl| 0 \bigl\rangle\bigl|$ with fixed $(\lambda,\lambda')$ chosen between the $\mathcal{W}_2^e$ and $\mathcal{W}_2^p$ phases, where $\beta_k^\dag(\vec{\lambda})$ is the $k$-th mode lower band creation operator with model parameter $\vec{\lambda}$. Note that $\mathcal{F}=\prod_k\mathcal{F}_{k}$ \cite{zeng2024}, in turn the zeros pinned at certain $k$-modes necessitate $\mathcal{F}$ to be zero. The pinning of such zeros are also present when one considers two GSs from phases with different winding numbers, which accounts for the zeros of the fidelity observed in the $\mathcal{W}_1^e$, $\mathcal{W}_0^e$, and $\mathcal{W}_{-1}^{e/p}$ regimes in Fig. \ref{fig:fidelitymap}(a), consistent with Ref. \cite{huang2016,zeng2024,sacramento2019}. On the contrary, when one considers $\mathcal{F}_{k}$ between the same phase, $\nexists \ \mathcal{F}_{k}=0$ [see coral curve in Fig. \ref{fig:fidelitymap}(b)]. Therefore, $\mathcal{W}_2^e$ and $\mathcal{W}_2^p$ are two distinct phases. Similar arguments hold true for other winding numbers. Additional evidences for $\mathcal{W}_m^e$ and $\mathcal{W}_m^p$ to be different phases and the nomenclature is clarified in the SM \cite{evidence_nomen_sm}.

\emph{Real-space configuration of the GS in the ESSH model.---} Here, we unveil the real-space configuration of the GS of different winding number phases in the ESSH model, which is not a priori obvious. Figure \ref{fig:fidelitymap}(c-f) show the GS configuration of the four winding number phases, where the $x$-axis is the integer corresponding to the binary representation of the Fock state, and the $y$-axis is the squared modulus of the corresponding weight ($|\xi_j|^2$). Most of the dominant Fock states are approximately equally weighted and they are shared by the same ``e'' and ``p'' phase.

After analyzing the dominant Fock states of the $\mathcal{W}_2^e$ phase in Fig. \mbox{\ref{fig:fidelitymap}}(e), one finds a generic pattern - for all cell $j$ in these Fock states, the sites $(j,B)$ and $(j+2,A)$ always have a pair of 0 and 1 (see SM for examples \cite{w2fock_sm}). Consequently, these sites favor the $t_d$ hopping. This is the physical intuition we gain - the GS of the $\mathcal{W}_2^e$ phase is approximately a superposition of all Fock states with the $t_d$ hopping present throughout the entire chain. This preference for $t_d$ hopping in the $\mathcal{W}_2^e$ phase is consistent with the behavior of entanglement entropies investigated in Ref. \cite{wong2024}. 
One can also perform a similar analysis on the $\mathcal{W}_0^e$ phase, and find that all the dominant Fock states in the $\mathcal{W}_0^e$ phase must possess a pair of 0 and 1 on the sites $(j,A)$ and $(j,B)$, for any $j$. More generally, we find a real space physical picture for any winding number phase of a general SSH model with arbitrary number of further neighbor hoppings:
\begin{quote}
\small
Consider $H = \sum_{j = 1}^{N} \sum_{i=\gamma}^{-\gamma}( t_i^{}c_{j,B}^\dagger c_{j+i,A}^{}  + \text{h.c.} )$. If $t_m\gg t_l, \ \forall \ \ l\neq m$, then $\mathcal{W}=m$, and the GS consists of all Fock states that have pairs of 1 and 0 across all sites connected by $t_m$. 
\end{quote}

\begin{figure}
    \centering
    \includegraphics[width=8cm]{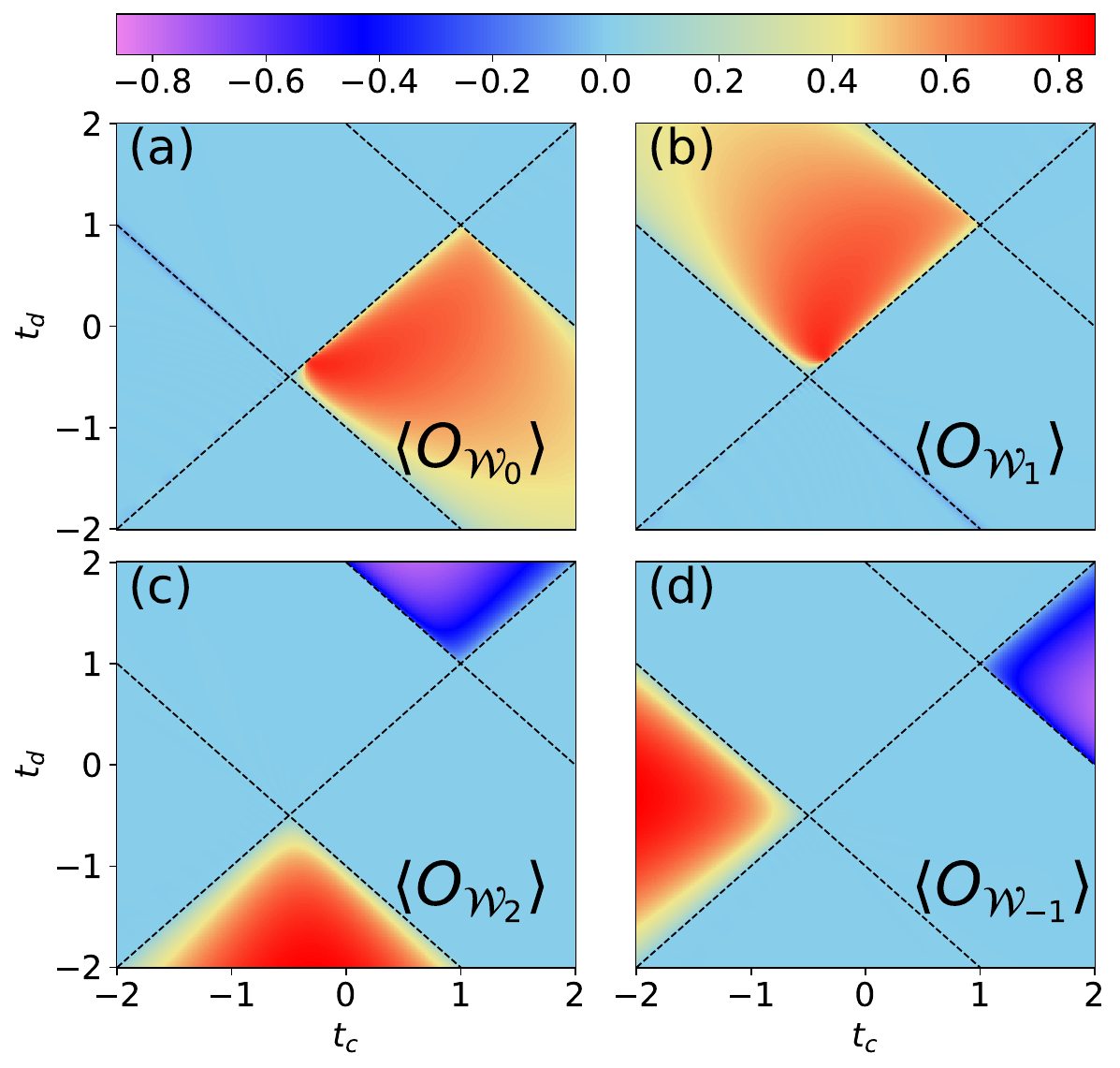}
    \caption{The GS expectation values of the OPs in Eq. (\ref{eqn:w0}-\ref{eqn:wn1}) \cite{gs_ev_derivation_sm}. Here $t_a=t_b=-1$, $N=200$, and $n=\frac{N}{2}
    - 1=99$. Black dashed lines indicate the phase boundaries determined from the winding numbers. 
    }
    \label{fig:ssh_OP}
\end{figure}

\emph{OPs in the ESSH model.---}The above insights are critical for the following discussion. For finding an OP for Eq. (\ref{eqn:H_SSH}), the target is to find a set of OPs $\{O_{\mathcal{W}_0} ,O_{\mathcal{W}_1} ,O_{\mathcal{W}_2} ,O_{\mathcal{W}_{-1}}\}$ such that, in its own phase, the corresponding OP has the largest expectation value among the others. 
For instance, we require $\langle O_{\mathcal{W}_2}\rangle$ takes the largest value in regions whenever $\mathcal{W}=2$, when comparing with $\langle O_{\mathcal{W}_0}\rangle$, $\langle O_{\mathcal{W}_1}\rangle$ and $\langle O_{\mathcal{W}_{-1}}\rangle$. A concrete example is shown in Fig. \ref{fig:ssh_OP}. 

Motivated by the physical observation above and consider an example of $N=8$, we write:
\begin{align}
O_{\mathcal{W}_2}^{(3)}&=c^\dag_{j,B}c_{j+2,A}^{}c^\dag_{j+1,B}c_{j+3,A}^{}c^\dag_{j+2,B}c_{j+4,A}^{}\ ,\\
&=c^\dag_{1B}c_{3A}^{}c^\dag_{2B}c_{4A}^{}c^\dag_{3B}c_{5A}^{} \ \ \ \ \ \ \text{if }\  j=1,
\label{eqn:W2_3}
\end{align}
where the superscript denotes the presence of 3 hopping terms in this OP. The value of $j$ does not matter, since PBC is imposed. The reason for including at least three hopping terms will be clarified later. Here we only include three hopping terms in the proposed OP for the sake of demonstrating how the scheme works. 

We note that any Fock state that does not satisfy the following form vanishes automatically when acted on by the operator in Eq. (\ref{eqn:W2_3}):
\begin{align}
    |\widehat{n_{1A}^{}\ 0}\ \ \widehat{n_{2A}^{}\ 0}\ \ \widehat{1^{}_{\ }\ 0^{}_{\  }}\ \ \widehat{1\ n_{4B}^{}}\ \ \widehat{1\ n_{5B}^{}} \  \cdots \  \widehat{n_{8A}^{}\ n_{8B}^{}}\rangle,
    \label{eqn:fockw2}
\end{align}
where the hats denote a unit cell, and the unspecified site occupancies can be either 0 or 1. An example of such is $|\widehat{0 0}\ \widehat{1 0}\ \widehat{1 0}\ \widehat{1 1}\ \widehat{1 1}\ \widehat{0 1}\ \widehat{0 1}\ \widehat{0 0}\rangle$. The Fock state in Eq. (\ref{eqn:fockw2}) has more significant weight in the GS of the $\mathcal{W}_2^{e/p}$ phase, and is less weighted in the GSs of other phases with different winding numbers. The reason is as follows: Focusing on the cells $j=1,2,3,4,5$, the Fock state in Eq. (\ref{eqn:fockw2}) satisfies one of the dominant configurations in the GS of the $\mathcal{W}_2^e$ phase automatically. However, there is only $\frac{1}{16}$ probability that it belongs to the GS of the $\mathcal{W}_0^e$ phase, likewise for the $\mathcal{W}_{-1}^e$ phase, and a probability of $\frac{1}{4}$ for the GS of the $\mathcal{W}_1^e$ phase. The key insight here is whenever we act the operator $O_{\mathcal{W}_2}^{(3)}=c^\dag_{1,B}c_{3,A}^{}c^\dag_{2,B}c_{4,A}^{}c^\dag_{3,B}c_{5,A}^{}$ onto a state that is not the GS of $\mathcal{W}_2^e$, it will give a very small value, compared to a GS that belongs to $\mathcal{W}_2^e$. The operator in Eq.  (\ref{eqn:W2_3}) effectively screens out all the Fock states that have no configuration associated with the $t_d$ hopping. Therefore, the more number of $t_d$ hopping terms the OP includes, the more accurate it should describe the $\mathcal{W}_2^e$ phase. 

For the case of $\mathcal{W}_0^e$/$\mathcal{W}_1^e$/$\mathcal{W}_{-1}^e$, one can repeat similar procedure illustrated above, and find out that the dominant Fock states also have the 0 and 1 pairs between the corresponding sites associated with $t_a$/$t_b$/$t_c$ hoppings, analogous to the relation between $\mathcal{W}_2^e$ and $t_d$ hoppings. Thus, we have 
\begin{align}
\label{eqn:w0_3}
O_{\mathcal{W}_0}^{(3)}=c^\dag_{1,A}c_{1,B}^{}c^\dag_{2,A}c_{2,B}^{}c^\dag_{3,A}c_{3,B}^{}\ ,\\
\label{eqn:w1_3}
O_{\mathcal{W}_1}^{(3)}=c^\dag_{1,B}c_{2,A}^{}c^\dag_{2,B}c_{3,A}^{}c^\dag_{3,B}c_{4,A}^{}\ ,\\
\label{eqn:w-1_3}
O_{\mathcal{W}_{-1}}^{(3)}=c^\dag_{1,A}c_{2,B}^{}c^\dag_{2,A}c_{3,B}^{}c^\dag_{3,A}c_{4,B}^{}\ ,
\end{align}
analogous to Eq. (\ref{eqn:W2_3}). 

We can promote the potential OPs to include, instead of just three, to $n$ hopping terms, namely,
\begin{align}
\label{eqn:w0}
O_{\mathcal{W}_0} &= 2^{n-1}\left(\prod_{j=1}^n c^\dag_{j,B}c_{j,A}^{} + \text{h.c.}\right) ,\\
\label{eqn:w1}
O_{\mathcal{W}_1} &= 2^{n-1} \left(\prod_{j=1}^n c^\dag_{j+1,A}c_{j,B}^{} + \text{h.c.}\right) ,\\
\label{eqn:w2}
O_{\mathcal{W}_2}&= 2^{n-1} \left(\prod_{j=1}^n c^\dag_{j+2,A}c_{j,B}^{} + \text{h.c.}\right) ,\\
\label{eqn:wn1}
O_{\mathcal{W}_{-1}} &= 2^{n-1} \left(\prod_{j=1}^n c^\dag_{j+1,B}c_{j,A}^{} + \text{h.c.}\right).
\end{align}
This is the OP for Eq. (\ref{eqn:H_SSH}), and it turns out that $n=\frac{N}{2}\pm 1$ gives the best result,  which is showcased in Fig. \ref{fig:ssh_OP}. Additionally, the OPs distinguish the $\mathcal{W}_m^e$ and $\mathcal{W}_m^p$ phases by its sign, which does more than the traditional winding number. Furthermore, the magnitude of the OP indicates the depth of the phase, information not captured by the winding number. 

If one considers the general SSH model $H = \sum_{j = 1}^{N} \sum_{i=\gamma}^{-\gamma}( t_i^{}c_{j,B}^\dagger c_{j+i,A}^{}  + \text{h.c.} )$, the OPs generalize to:
\begin{align}
\label{eqn:generalise_ssh_op}
O_{\mathcal{W}_m}&=  2^{n-1} \left( \prod_{j=1}^n c^\dag_{j+m,A}c_{j,B}^{} + \text{h.c.}\right)\  \text{for $m\in \mathbb{Z}$},
\end{align}
The result is shown in the SM \cite{generalizeSSH_sm}.

\begin{figure}
    \centering
    \includegraphics[width=8cm]{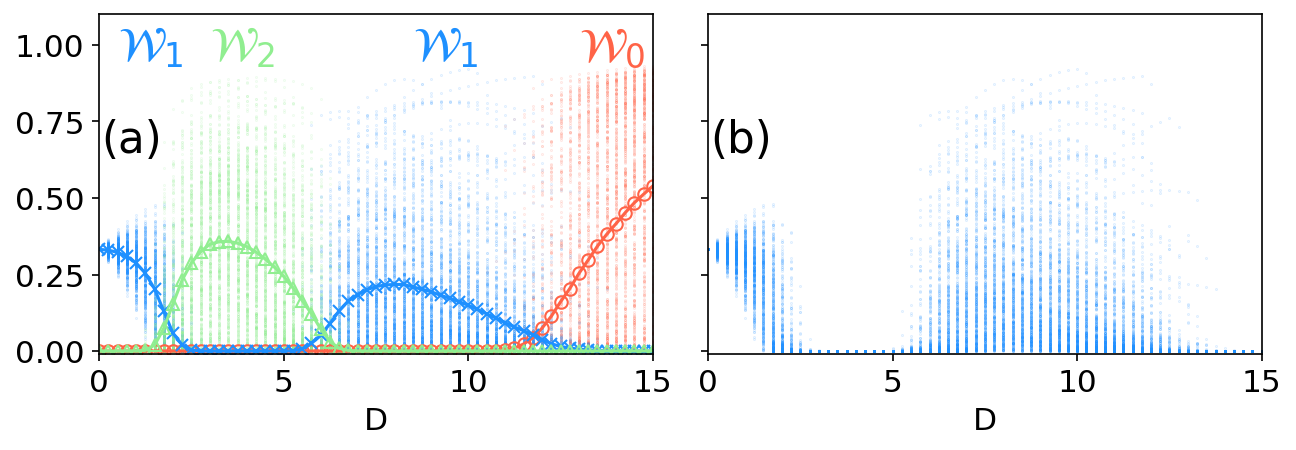}
    \caption{$|\langle O_{\mathcal{W}_2}\rangle|$ (green), $|\langle O_{\mathcal{W}_1}\rangle|$ (blue), and $|\langle O_{\mathcal{W}_0}\rangle|$ (red) as a function of $D$ in the disordered SSH model in Eq. (\ref{eqn:H_disorder}). Dots represent $|\langle O_{\mathcal{W}_m}\rangle|$, for $m\in\{0,1,2\}$, of each of the 400 realizations for each $D$ value, and the markers with solid lines are the sample averages of $|\langle O_{\mathcal{W}_m}\rangle|$. The chosen parameters is $(t_a'',t_b'',t_d'')=(1,-0.8,-1.74)$, and $(D_a,D_b)=(D,0)$ (same setting as in Ref. \cite{Hsu2020}). (b) is a replicate of (a), except it only shows the measured $|\langle O_{\mathcal{W}_1}\rangle|$ on each of the samples. Here, $D_d=0$, $N=1000$, and $n=\frac{N}{2}-1=499$.}
    \label{fig:disorder}
\end{figure}

\emph{Disordered SSH model.---}The Hamiltonian reads \cite{song2014,lin2021,Hsu2020}:
\begin{align}
    H_{\text{disorder}} &= \sum_{j = 1}^{N} ( t_{a,j}''c_{j,A}^\dagger c_{j,B}^{} + t_{b,j}''c_{j,B}^\dagger c_{j + 1,A}^{} \nonumber\\&\qquad + t_{d,j}''c_{j,B}^\dagger c_{j + 2,A}^{} + \text{h.c.} ),
    \label{eqn:H_disorder}
\end{align}
where the hopping parameters $(t_{a,j}'',t_{b,j}'',t_{d,j}'')$ are site-dependent, and we prohibit the $t_c$-type of hoppings for simplicity. We set $t_{m,j}''=t_m''+D_m\epsilon_{m,j}$, where $m\in\{a,b,d\}$ and $\epsilon_{m,j}\in(-\frac{1}{2},\frac{1}{2})$ is a random number sampled from a uniform distribution, and $D_m$ is the disorder strength for the respective type of hoppings. To calculate the many-body GS expectation values of the OPs in Eqs. (\ref{eqn:w0}-\ref{eqn:w2}), $\langle O\rangle = \sum_i f_i \langle \psi_i|O|\psi_i\rangle$ is used, where $f_i$ is the Fermi-Dirac distribution and $|\psi_i\rangle$ is the $i$-th eigenstate of the single-particle Hamiltonian in Eq. (\ref{eqn:H_disorder}).  

Figure \ref{fig:disorder}(a) shows our OPs in Eq. (\ref{eqn:w0}-\ref{eqn:w2}) for $n=\frac{N}{2}-1$ as a function of $D$. The ensemble averages over 400 disorder realizations, for each $D$. The crossings between our OPs match qualitatively well with the transitions reported in Ref. \cite{Hsu2020}, which are determined from only 50 disorder realizations using the non-commutative winding number \cite{shem2014}. We remark that such consistency can only be reached in large systems beyond those accessible by exact diagonalization.

Furthermore, the measured OPs on each of the disorder realizations attain significant finite values in their own phase and vanish elsewhere. Figure \ref{fig:disorder}(b) shows the $|\langle O_{\mathcal{W}_1}\rangle|$ of all the random samples, using the same parameters as Fig. \ref{fig:disorder}(a). In addition, our OPs convey how ``deep" the system is in a phase via their magnitudes, which lack from the non-commutative winding number, similar to the equilibrium case in Fig. \ref{fig:ssh_OP}.

\begin{figure}
    \centering
    \includegraphics[width=8cm]{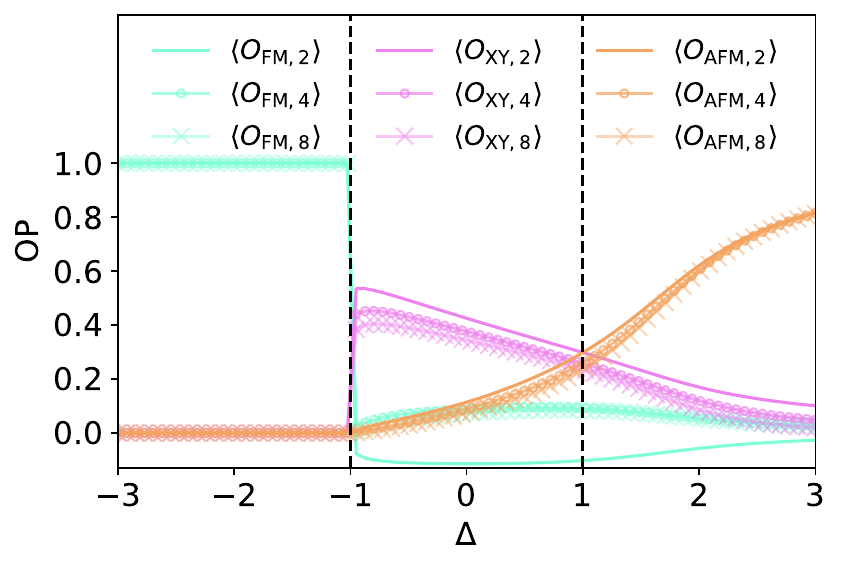}
    \caption{The OP in Eqs. (\ref{op:fmn}-\ref{op:xyn}) with $n=2,4,8$ and $D=4$  as a function of the model parameter $\Delta$. Here, a system of 20 spins is considered. Vertical dashed lines indicate the transition points.}
    \label{fig:xxzop}
\end{figure}

\emph{OPs of the XXZ model.---}The Hamiltonian reads
$H_{\text{XXZ}}= \sum_j\bigl(  \sigma_j^x\sigma_{j+1}^x+\sigma_j^y\sigma_{j+1}^y+\Delta\sigma_j^z\sigma_{j+1}^z\bigl)$. The model exhibits two critical points: $\Delta=- 1$, where the transition is first order; $\Delta=1$, where the transition is of BKT type \cite{justino2012}. There are three phases that can exist in this model: (1) $\Delta<-1$ is the ferromagnetic (FM) phase, where the system favors all spins aligning upward or downward; (2) $-1<\Delta<1$ is known as the XY phase, where its order is less understood \cite{Cheraghi2019}; (3) $\Delta>1$ is anti-ferromagnetism (AFM), where anti-parallel alignment along a $z$ direction is preferred for the spins in the system. For a system of 16 spins, the GS of a deep AFM phase has two dominant Fock states:
\begin{align}
|1 0\ 1 0 \ 1 0 \ 1 0 \ 1 0 \ 1 0 \ 1 0 \ 1 0 \rangle,
\label{Fock:afm1010}\\
|0 1\ 0 1\ 0 1\ 0 1\ 0 1\ 0 1\ 0 1\ 0 1\rangle,
\label{Fock:afm0101}
\end{align}
where 1 and 0 refers to an up and a down spin, respectively. Separations between the sites in Eqs. (\ref{Fock:afm1010}-\ref{Fock:afm0101}) do not imply a unit cell, rather, they are just a visual aid.

For the FM phase, the lowest energy is two-fold degenerated in finite systems. The GS in the degenerated space is:
\begin{align}
    |\text{FM}\rangle &= \alpha|1 1\ 1 1  \cdots  1 1 \ 1 1 \rangle+\beta|0 0\ 0 0 \cdots 0 0 \ 0 0 \rangle,
    \label{Fock:fm}
\end{align}
where $|\alpha|^2+|\beta|^2=1$ and the equal sign in Eq. (\ref{Fock:fm}) is numerically exact for $\Delta<-1$.

For the XY phase, there are a large number of Fock states with comparable weights (see the SM \cite{xxz_sm}). The largest weighted Fock state is Eq. (\ref{Fock:afm1010}) and (\ref{Fock:afm0101}) (which is of the AFM configuration). For the other Fock states with smaller but comparable weights, their configurations contain some local violations to the AFM pattern in Eq. (\ref{Fock:afm1010}). To capture these local violations, spin exchange operator is needed. One suitable choice is 
\begin{align}
    O_{\text{XY}, 2}(i,D) &= \frac{1}{D}\sum_{j=1}^D(-1)^j\Bigl(\sigma^x_i\sigma^x_{i+j}+\sigma^y_i\sigma^y_{i+j} \Bigl)\nonumber\\
    &= \frac{1}{D}\sum_{j=1}^D\frac{(-1)^j}{2}\Bigl(\sigma^+_i\sigma^-_{i+j}+\sigma^-_i\sigma^+_{i+j} \Bigl),
\label{op:xy2}
\end{align}
It is easy to check $\langle O_{\rm{XY},2}\rangle$ is finite in the XY phase and almost vanishes in the other phases (see SM for details \cite{xxz_sm}). The simple structure in the GS of the AFM and FM phase also enable us to easily write down their OPs:
\begin{align}
\label{op:fm2}
    O_{\text{FM,2}}(i,D) &=\frac{1}{D}\sum_{j=1}^D\sigma^z_{i}\sigma^z_{i+j}\ , \\
    \label{op:afm2}
    O_{\text{AFM,2}}(i,D) &=\frac{1}{D}\sum_{j=1}^D(-1)^j\sigma^z_{i}\sigma^z_{i+j}\ . 
\end{align}

Figure \ref{fig:xxzop} shows the OPs in Eq. (\ref{op:xy2}), (\ref{op:fm2}) and (\ref{op:afm2}) as a function of $\Delta$ (the most opaque curves). The OP attains the highest values in its own phase and a smaller value in other phases, and the crossings with other OPs also correctly indicates the transitions, including the BKT one. 

It turns out a class of OPs (generalized from above):
\begin{align}
    O_{\text{FM},n}(i,D) &=\frac{1}{D}\sum_{j=1}^D\sigma^z_{i}\prod_{k=0}^{n-2}\sigma^z_{i+j+k}\ ,\label{op:fmn} \\
    O_{\text{AFM},n}(i,D) &=\frac{1}{D}\sum_{j=1}^D(-1)^{(j+1)}\sigma^z_{i}\prod_{k=0}^{n-2}\sigma^z_{i+j+k}\ ,    \label{op:afmn}\\
    O_{\text{XY},n}(i,D) &= \frac{1}{D}\sum_{j=1}^D(-1)^{(j+1)}
    \times\nonumber\\
    &\quad\Bigl(\sigma^x_i\prod_{k=0}^{n-2}\sigma^x_{i+j+k} +\sigma^y_i\prod_{k=0}^{n-2}\sigma^y_{i+j+k} \Bigl),\label{op:xyn}
\end{align}
are also capable of capturing its own phase accordingly and predicting the correct transitions for any $D\geqslant 2$ and $n\geqslant 2$, where $n$ is the number of $\sigma$ operators in the product of each term (see SM for details \cite{xxz_sm}). We show $n=4,8$ in Fig. \ref{fig:xxzop} by the more transparent curves. The correctness of detecting the BKT transition by above class of OPs is due to the isotropy of the model at $\Delta=1$.  

\emph{Conclusions.---}We demonstrate a general scheme to construct the OPs in board settings - the topological ESSH model and its disordered case, and the interacting spin-half XXZ model. 
In the ESSH model, we elucidate the real-space configuration of a winding number phase, which is not a priori obvious. Exploiting this result, we find the OPs for the generalized SSH model. Additionally, we theorize that, for each winding-number phase, there should be a $\mathcal{W}_m^e$ and $\mathcal{W}_m^p$ counterpart. This marks a fundamental shift in our understanding of the model. While our OPs are capable of distinguishing these two phases, their physical interpretation remains to be clarified in future work. The derived OPs are also applied to the disordered SSH model and successfully capture the phase transitions with further information on the depth of the phase.
Applying our scheme to the interacting XXZ model, we find a class of OPs that accurately predict the phase transitions, including the BKT transition, which is known to be difficult to detect in finite systems.

We acknowledge financial support from Research Grants Council  of  Hong  Kong  (Grant  No. CityU 11318722),  FCT under research unit: UID/04540: Center of
Physics and Engineering of Advanced Materials and
contract LA/P/0095/2020, LaPMET, Laboratory of Physics for Materials and
Emerging Technologies, and City University of Hong Kong (Grants No. 9610438, 7020156). PDS thanks the hospitality during his visit to City University of Hong Kong. 

\bibliography{bibfile}



\clearpage
\onecolumngrid




\section{Supplemental Material for
``Beyond Topological Invariants: Order Parameters from Dominant Fock-state Patterns"
}

\onecolumngrid

\subsection{Remarks on the scheme}
\label{app:method_sup}

\emph{Tools for determining critical points.---}If the phase boundary is unknown, various techniques can be used to estimate the phase transition point before applying our scheme. Examples include the entanglement entropy (and its derivative) \cite{yu2016,Osterloh2002,Osborne2002,Amico2008,Li2022}, fidelity \cite{gu2010,Zanardi2006,Quan2006} and its variants such as the fidelity susceptibility \cite{gu2010,You2007} and the fidelity map \cite{ali2021}, or the energy bond correlators \cite{yu2020,yang2022}. More recently, there were methods proposed utilizing the sign problem in quantum Monte Carlo \cite{ma2024,Mondaini2022} or combining correlations in configuration space in the Monte Carlo method \cite{Su2024} to locate the phase transitions. 

\emph{Useful strategies.---}We list some strategies in deducing the correct OPs according to our experience. First, when performing the analysis on the dominant Fock states, choose points that are deep inside the phase, if possible. Second, periodic boundary condition (PBC) is preferred to reduce the finite size effect. Third, if the dominant Fock states are very scattered, the OPs usually has the form of combinations of creation and annihilation operators (for fermionic models) or the raising and lowering operators (for spin models). On the other hand, if there are only few dominant Fock states in the GS, forming the OP with the number operators or the spin operators in the $z$ direction is more favorable. 
We will unveil more details in the following section when we demonstrate the scheme on finding the OPs.

\emph{Advantages.---}There are three advantages. First, the scheme helps to gain more physical intuition in the phase of interest, since we directly observe the dominant Fock states, which gives an immediate physical picture of the electronic \ok{or spin} configurations. Second, the form of the OP that can be constructed is in principle unrestricted, which makes our scheme versatile. \ok{Third, as compared to the reduced density matrix based schemes \cite{furukawa2006,gu2013}, our scheme does not have the ambiguity in choosing a suitable subsystem.} 



\begin{figure*}[h]
    \centering
    \includegraphics[width=16cm]{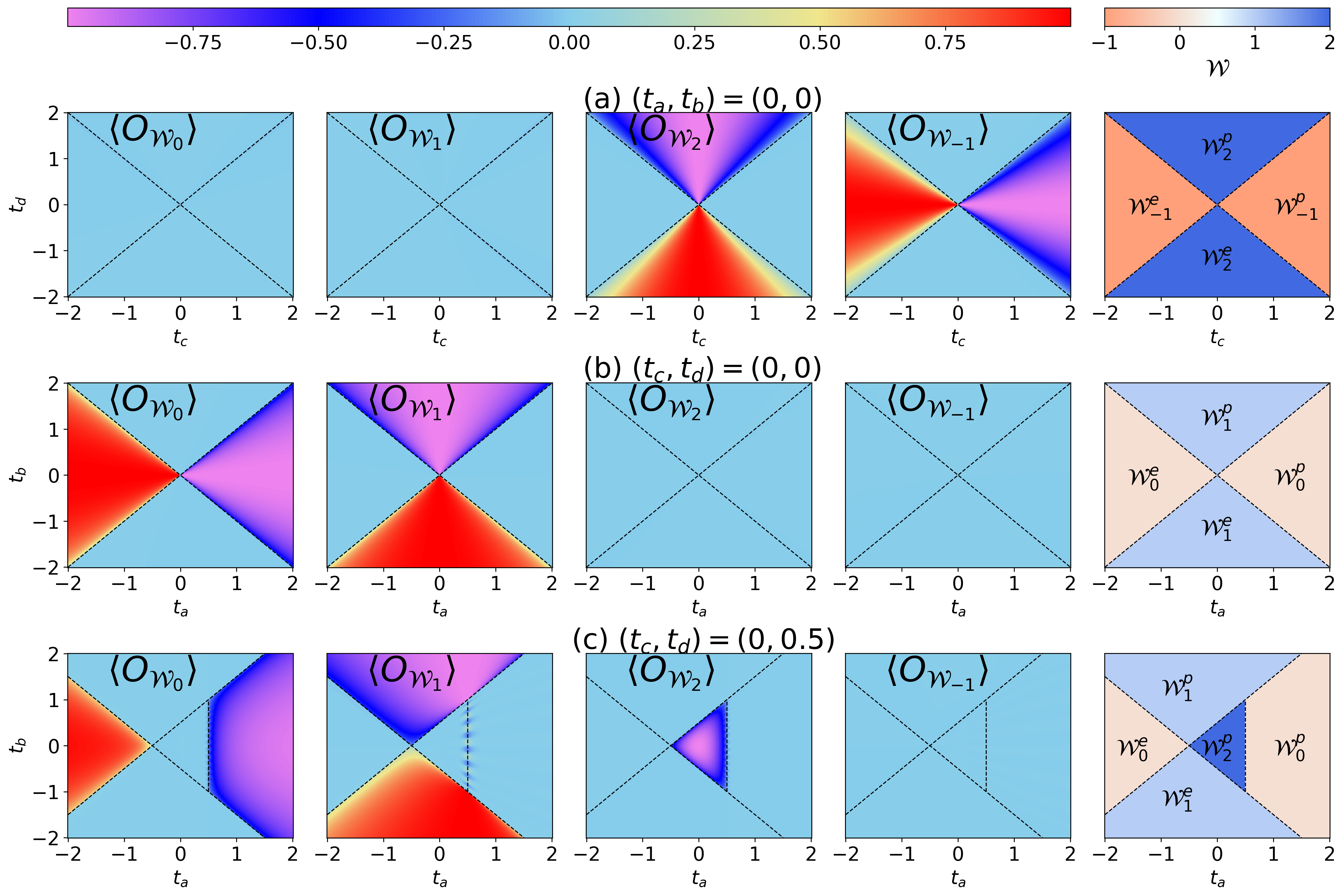}
    \caption{The four left columns show the ground-state expectation values of the OPs in Eq. (\ref{eqn:w0}-\ref{eqn:wn1}) with $N=200$ and $n=\frac{N}{2}-1$ for (a) $(t_a,t_b)=(0,0)$, as a function of $t_c$ and $t_d$; (b) $(t_c,t_d)=(0,0)$ and (c) $(t_c,t_d)=(0,0.5)$, both as a function of $t_a$ and $t_b$. The rightmost column shows the phase diagram. Black dashed lines indicate the phase boundaries determined from the winding numbers. }
    \label{fig:w0w1_w2w-1}
\end{figure*}

\subsection{Anderson Orthogonality Catastrophe}
\label{app:ortho_catastrophe}

As the system size increases, the non-zero fidelity region shrinks [Fig. \ref{fig:fidelitymap}(a)]. This is due to Anderson orthogonality catastrophe \cite{anderson1967}. The catastrophe implies that, even when the two parameter points are arbitrarily close and lie within the same phase, the ground-state fidelity approaches zero in the thermodynamic limit \cite{zeng2024}. To circumvent this, we consider $\mathcal{F}_k(\vec\lambda,\vec\lambda')$ as a function of $k$ to confirm if the fidelity is finite in the main text. As shown in Fig. \ref{fig:fidelitymap}(b), there is no pinning of zeros for the mode-resolved fidelity when $\vec\lambda$ and $\vec\lambda'$ both lie inside the $\mathcal{W}_2^e$ phase. This is also true in general for any two GSs from the same phase (i.e. the phase with identical winding number and superscript).

\subsection{Additional evidences for insufficiency of the winding number}
\label{app:twoMoreEvidences}

Besides the analysis of the fidelity presented in the main text, we provide two further evidence that $\mathcal{W}_2^e$ and $\mathcal{W}_2^p$ are truly two different phases.

\emph{Evidence 2: The existence of gap closing point between the $\mathcal{W}_m^e$ and $\mathcal{W}_m^p$ phases.---}Consider the $t_c-t_d$ phase diagram with $(t_a,t_b)=(-1,-1)$ in Fig. \ref{fig:fidelitymap}(a). If we vary $(t_a,t_b)$ from $(-1,-1)$ to $(0,0)$, the $\mathcal{W}_1^e$ and $\mathcal{W}_0^e$ phases fade away as shown in the rightmost plot in Fig. \ref{fig:w0w1_w2w-1}(a). Focusing on this phase diagram, one also notice that $(t_c,t_d)=(0,0)$ is a gapless point where a quantum phase transition takes place. There is a quantum phase transition between the $\mathcal{W}_2^e$ and $\mathcal{W}_2^p$ phases, as well as between the $\mathcal{W}_{-1}^e$ and $\mathcal{W}_{-1}^p$ phases. This strongly suggests that phases with the ``e" and ``p" labels are distinct. In fact, $\mathcal{W}_1^e$($\mathcal{W}_0^e$) also has a counterpart phase, namely $\mathcal{W}_1^p$($\mathcal{W}_0^p$) [see the rightmost plot in Fig. \ref{fig:w0w1_w2w-1}(b)]. 

\begin{figure}
    \centering
    \includegraphics[width=8cm]{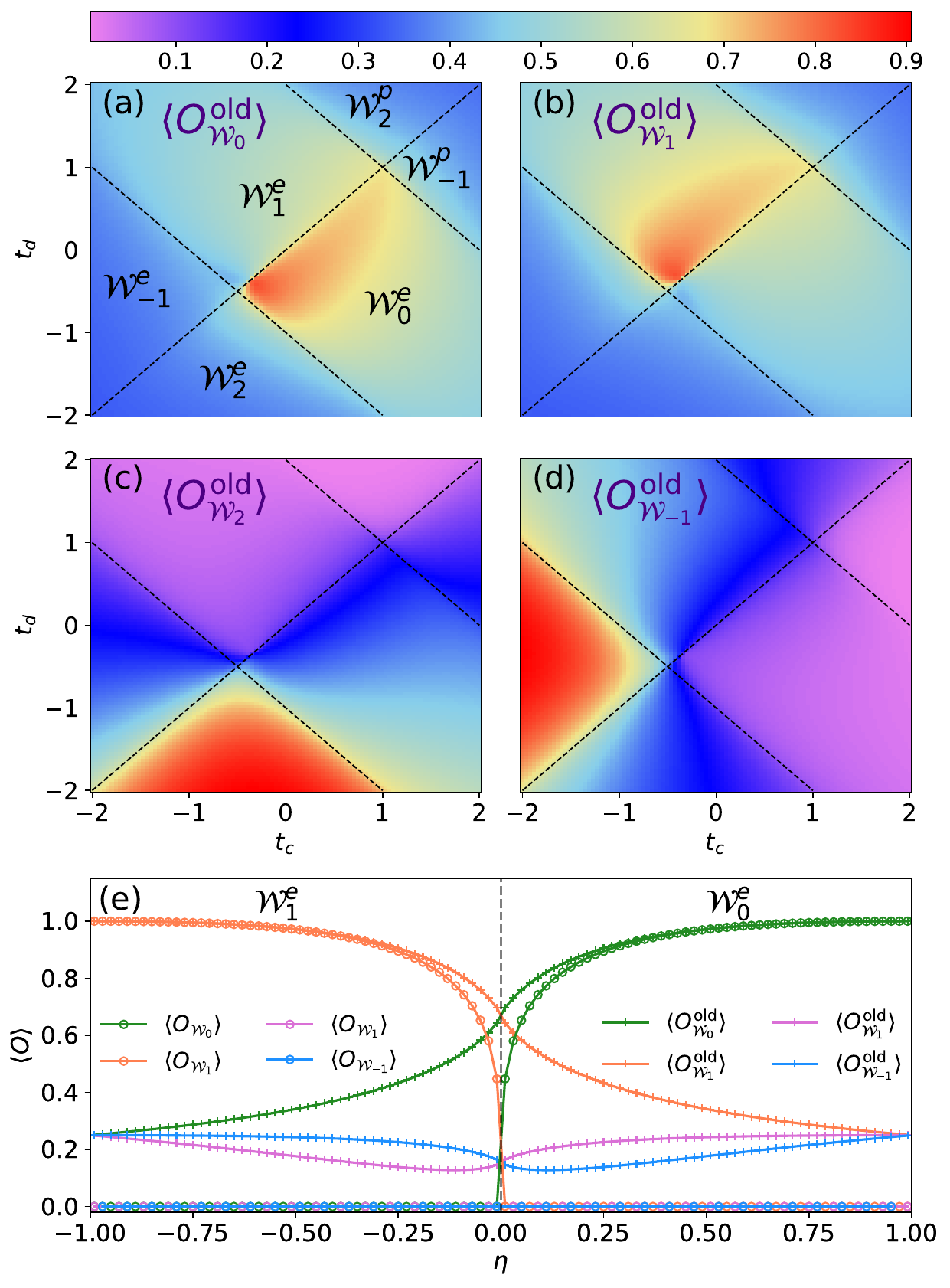}
    \caption{The OPs in Eqs. (\ref{eqn:gu2013_0}-\ref{eqn:gu2013_n1}) constructed from the scheme in Ref.\cite{gu2013,yu2019} as a function of $t_c$ and $t_d$ with $(t_a,t_b)=(-1,-1)$ for the (a) $\mathcal W_0^e$, (b) $\mathcal W_1^e$, (c) $\mathcal W_2^e$, and (d) $\mathcal W_{-1}^e$ phases. Black dashed lines indicate the phase boundaries determined by the winding numbers. Plot (e) compares the OPs in Eqs.(\ref{eqn:gu2013_0}-\ref{eqn:gu2013_n1}) with those in Eqs. (\ref{eqn:w0}-\ref{eqn:wn1}) in the conventional SSH model, i.e.  $t_c=t_d=0$ and $t_a=-(1+\eta)$ and $t_b=-(1-\eta)$. The vertical dashed line indicates the topological phase transition at $\eta=0$. Both the expectation values of $O_{\mathcal{W}_{2}}$ and $O_{\mathcal{W}_{-1}}$ are zero for all $\eta$. 
    System of $N=200$ is considered in all the plots here. }
    \label{fig:2016op}
\end{figure}

\emph{Evidence 3: Opposite behavior of physical quantities between $\mathcal{W}_m^e$ and $\mathcal{W}_m^p$ phases.---}In addition, the $\mathcal{W}_m^p$ and $\mathcal{W}_m^e$ phases exhibit distinct behavior in real-space, reinforcing the necessity for classifying them as different phases. Consider the OPs found in previous work \cite{gu2013,yu2016,yu2019}, which reads
\begin{align}
     \label{eqn:gu2013_0}
    O_{\mathcal{W}_0}^{\text{old}} = \frac{1}{2}&(c^\dag_{j,A}c^{}_{j,B}+\text{h.c.})-n^{}_{j,B}n^{}_{j,A}+\frac{1}{2}(n^{}_{j,B}+n^{}_{j,A}),\\ 
    \label{eqn:gu2013_1}
    O_{\mathcal{W}_{1}}^{\text{old}} =  \frac{1}{2}&(c^\dag_{j+1,A}c^{}_{j,B}+\text{h.c.})-n^{}_{j,B}n^{}_{j+1,A}+\frac{1}{2}(n^{}_{j,B}+n^{}_{j+1,A}),
\end{align}
and that for the $\mathcal{W}_2$ and $\mathcal{W}_{-1}$ phases derived following the same scheme
\begin{align}
\label{eqn:gu2013_2}
    O_{\mathcal{W}_2}^{\text{old}} = \frac{1}{2}&(c^\dag_{j+2,A}c^{}_{j,B}+\text{h.c.})-n^{}_{j,B}n^{}_{j+2,A}+\frac{1}{2}(n^{}_{j,B}+n^{}_{j+2,A}), \\
    \label{eqn:gu2013_n1}
    O_{\mathcal{W}_{-1}}^{\text{old}} = \frac{1}{2}&(c^\dag_{j,A}c^{}_{j+1,B}+\text{h.c.})-n^{}_{j+1,B}n^{}_{j,A}+\frac{1}{2}(n^{}_{j+1,B}+n^{}_{j,A}).
\end{align}
These OPs work well in the original SSH model, i.e. $t_c=t_d=0$. As shown in Fig. \ref{fig:2016op}(e),  the GS expectation value of the OPs in Eqs. (\ref{eqn:gu2013_0}-\ref{eqn:gu2013_n1}) is the largest in their respective phase. 
However, these OPs become inadequate when further neighbor hoppings ($t_c, t_d\ne 0$) are introduced. Figure \ref{fig:2016op}(a-d) displays the GS expectation value the OPs in Eqs. (\ref{eqn:gu2013_0}-\ref{eqn:gu2013_n1}) for $t_a=t_b=-1$ as a function of $t_c$ and $t_d$. Certain parameter regimes are mispredicted by these order parameters. For example, in the $\mathcal{W}_2^e$ phase, $\langle O_{\mathcal{W}_0}^{\text{old}}\rangle$ and $\langle O_{\mathcal{W}_1}^{\text{old}}\rangle$ attain relatively large values compared with the others, and one would incorrectly label the region as the $\mathcal{W}_0$ or the $\mathcal{W}_1$ phase. Nevertheless, $\langle O_{\mathcal{W}_2}^{\text{old}}\rangle$ attains a large value in the $\mathcal{W}_2^e$ phase but almost zero in the $\mathcal{W}_2^p$ phase further supports that these phases are different. 

The hollow circles in Fig. \ref{fig:2016op}(e) also shows the GS expectation value of the OPs in Eqs. (\ref{eqn:w0}-\ref{eqn:wn1})  derived in this work using the scheme introduced above. The results showcase the expected vanishing behavior of an OP outside its own respective phase. For example, $\langle O_{\mathcal{W}_0}\rangle $ ($\langle O_{\mathcal{W}_1}\rangle $) is zero in the $\mathcal{W}_1^e$ ($\mathcal{W}_0^e$) phase. Similarly, $\langle O_{\mathcal{W}_2}\rangle $ and $\langle O_{\mathcal{W}_{-1}}\rangle $ are always zero in the whole region. Furthermore, when one enters the $\mathcal{W}_1^e$ ($\mathcal{W}_0^e$) phase from $\mathcal{W}_0^e$ ($\mathcal{W}_1^e$),  $\langle O_{\mathcal{W}_0}\rangle $ ($\langle O_{\mathcal{W}_1}\rangle $)  drops from a finite value to zero abruptly.  Such a sharp change not only appears in this original SSH model, but also in the extended SSH model. It also holds true for the generalized version of the SSH model, where an arbitrary number of further neighbor hoppings and winding numbers are present (Fig. \ref{fig:generalise}).

\subsection{Nomenclature of the $\mathcal{W}_m^e$ and $\mathcal{W}_m^p$ phase}
Here, we remark the reason behind the nomenclature of the $\mathcal{W}_m^e$ and $\mathcal{W}_m^p$ phase. ``e'' carries the meaning of electron-like, while ``p'' has the meaning of hole-like. 
Typically, the hopping integral should be strictly negative in the tight binding model. However, if one considers hoppings to be positive, it can be thought of as having a negative effective mass, which is the defining property for a hole. Hole-like phases emerge when a particular positive hopping dominates. Therefore, we call it the hole-like phase, with some winding number. Same reasons apply to the electron-like phase. One should not confuse with the hole-like and electron-like phase in relation to the Lifshitz transition in the literature, in which the nomenclature in the later case is based on the topology of the Fermi surface \cite{Volovik2018,Wu2018,Chen2012,Sakai2009}.

\subsection{Dominant Fock states of the $W_2^{p/e}$ phase in the ESSH Model}
\label{app:ssh_OP}
Some examples of the dominant Fock states in Figure \ref{fig:fidelitymap}(e) are as follows:
\begin{equation}
\begin{split}
&|\widehat{1 1}\ \widehat{1 1}\ \widehat{0 1}\ \widehat{0 1}\ \widehat{0 1}\ \widehat{0 1}\ \widehat{0 0}\ \widehat{0 0}\rangle, \ 
|\widehat{1 1}\ \widehat{1 1}\ \widehat{0 1}\ \widehat{0 1}\ \widehat{0 1}\ \widehat{0 0}\ \widehat{0 0}\ \widehat{1 0}\rangle, \ 
|\widehat{1 1}\ \widehat{1 1}\ \widehat{0 1}\ \widehat{0 1}\ \widehat{0 0}\ \widehat{0 0}\ \widehat{1 0}\ \widehat{1 0}\rangle, \ 
|\widehat{1 1}\ \widehat{1 1}\ \widehat{0 1}\ \widehat{0 0}\ \widehat{0 0}\ \widehat{1 0}\ \widehat{1 0}\ \widehat{1 0}\rangle, \\
&|\widehat{1 1}\ \widehat{1 1}\ \widehat{0 0}\ \widehat{0 0}\ \widehat{1 0}\ \widehat{1 0}\ \widehat{1 0}\ \widehat{1 0}\rangle, \ 
|\widehat{1 1}\ \widehat{1 0}\ \widehat{0 0}\ \widehat{1 0}\ \widehat{1 0}\ \widehat{1 0}\ \widehat{1 0}\ \widehat{1 0}\rangle, \ 
|\widehat{1 1}\ \widehat{0 1}\ \widehat{0 1}\ \widehat{0 1}\ \widehat{0 1}\ \widehat{0 1}\ \widehat{0 0}\ \widehat{0 1}\rangle, \ 
|\widehat{1 1}\ \widehat{0 1}\ \widehat{0 1}\ \widehat{0 1}\ \widehat{0 1}\ \widehat{0 0}\ \widehat{0 0}\ \widehat{1 1}\rangle, \\
&|\widehat{1 1}\ \widehat{0 1}\ \widehat{0 1}\ \widehat{0 1}\ \widehat{0 0}\ \widehat{0 0}\ \widehat{1 0}\ \widehat{1 1}\rangle, \
|\widehat{1 1}\ \widehat{0 1}\ \widehat{0 1}\ \widehat{0 0}\ \widehat{0 0}\ \widehat{1 0}\ \widehat{1 0}\ \widehat{1 1}\rangle, \
|\widehat{0 0}\ \widehat{1 0}\ \widehat{1 0}\ \widehat{1 1}\ \widehat{1 1}\ \widehat{0 1}\ \widehat{0 1}\ \widehat{0 0}\rangle, \ \cdots  
\end{split}
\label{eqn:list_of_w2_fock}
\end{equation}
They have almost the same weight (i.e. $|\xi_j|^2$) if we consider the GS deep inside the $\mathcal{W}_2^e$ phase. All the Fock states above are expressed in the form of $|\widehat{n_{1,A}^{} n_{1,B}^{}}\ \ \widehat{n_{2,A}^{} n_{2,B}^{}} \ \cdots \  \widehat{n_{8,A}^{} n_{8,B}^{}}\rangle$ and the hat above denotes a unit cell, 0 and 1 indicate the occupancy at each site. We remark that these dominant Fock states are shared by the same ``e'' and ``p'' phase, e.g. $\mathcal{W}_2^e$ and the $\mathcal{W}_2^p$ phases. However, their relative signs of $\xi_j$ can be different.

\subsection{Derivation of the ground state expectation value of the OPs of the extended SSH model}
\label{app:2-pointCorrelation}

The model in Eq. (\ref{eqn:H_SSH}) can be solved exactly by performing the Fourier transformation
\begin{equation}\label{eqn:FT_rule}
    \begin{split}
    a_k &= \frac{1}{\sqrt{N}} \sum_j e^{-ikj} c_{j,A} \ ,\\
    b_k &= \frac{1}{\sqrt{N}} \sum_j e^{-ikj} c_{j,B}\ ,
    \end{split}
\end{equation}
which enable us to rewrite the Hamiltonian into the following form: 
\begin{equation}
    H_{\text{SSH}} = \sum_k 
    \begin{pmatrix}
        a_k^\dagger &b_k^\dagger 
    \end{pmatrix}
    H_k
    \begin{pmatrix}
        a_k\\b_k
    \end{pmatrix},
    \label{eqn:H_k}
\end{equation}
where 
$H_k= \begin{pmatrix}
0 & f_k\\
f^*_k & 0\end{pmatrix}$ 
is the Hamiltonian in the k-space representation, with $f_k=t_a+t_be^{-ik}+t_ce^{ik}+t_de^{-2ik}$.         
Each $H_k$ in Eq. (\ref{eqn:H_k}) can be diagonlized by a unitary matrix $U$ satisfying $UU^\dag=U^\dag U =I$. We can write 
\begin{align}
H_k&=\begin{pmatrix}
    a_k^\dagger &b_k^\dagger 
\end{pmatrix} UU^\dag 
\begin{pmatrix}
    0 & f_k\\
    f^*_k & 0
\end{pmatrix}UU^\dag     
\begin{pmatrix}
    a_k\\b_k
\end{pmatrix}= 
\begin{pmatrix}
    \alpha_k^\dagger &\beta_k^\dagger 
\end{pmatrix}\Lambda_k
\begin{pmatrix}
    \alpha_k \\\beta_k 
\end{pmatrix},
\end{align}
where $\Lambda_k$ is diagonal. A straightforward diagonalisation shows that $U=\frac{1}{\sqrt{2}}
\begin{pmatrix}
    \frac{f_k}{E_k} & -\frac{f_k}{E_k}\\
    1 &1
\end{pmatrix}$, where $E_k=|f_k|$, and
\begin{equation}
\begin{split}
    \alpha_k &= \frac{1}{\sqrt{2}}\Bigl( \frac{f_k^*}{E_k}a_k + b_k\Bigl),\\
    \beta_k &= \frac{1}{\sqrt{2}}\Bigl( -\frac{f_k^*}{E_k}a_k + b_k\Bigl).
\end{split}    
\end{equation}
Therefore, 
\begin{align}
    H_k = E_k\alpha^\dag_k\alpha^{}_k - E_k\beta^\dag_k\beta^{}_k.
\end{align}
Here, $\alpha_k^\dag$ ($\beta_k^\dag$) creates a quasi-particle in the $k$-mode with energy $E_k$ ($-E_k$) in the upper (lower) band. 
At half-filling, the many-body ground-state is obtained by filling the lower band, i.e.
\begin{align}
    |\text{GS}\rangle = \prod_{k\in 1^\text{st}\text{B.Z.}}\beta_k^\dag|0\rangle. 
    \label{eqn:GS}
\end{align}
Using Eq. (\ref{eqn:FT_rule}), the real-space annihilation operators can be expressed as:
\begin{equation}\label{eqn:cjA_cjB}
\begin{split}
    c_{j,A}^{} &=  \frac{1}{\sqrt{2N}} \sum_k e^{ikj}\frac{E_k}{f_k^*} \Bigl( \alpha_k^{} - \beta_k^{} \Bigl),\\
    c_{j,B}^{} &=  \frac{1}{\sqrt{2N}} \sum_k e^{ikj} \Bigl( \alpha_k^{} + \beta_k^{} \Bigl).
\end{split}    
\end{equation}
Utilizing Eq. (\ref{eqn:cjA_cjB}), one obtains
\begin{align}
    \langle c_{j,B}^\dag(t)
c_{j+r,A}^{}(t)\rangle = \frac{-1}{2N}\sum_{k}e^{ikr}\frac{E_k}{f_k^{*}}.
\label{eqn:ciB(t)cjA(t)}
\end{align}
The OPs in Eqs. (\ref{eqn:w0}-\ref{eqn:generalise_ssh_op}) can be expressed in form of the determinant of a Toeplitz matrix whose elements are two-point correlation functions \cite{Schonhammer2017}, i.e.
\begin{align}
    &\quad \langle \prod_{j=1}^{n} c_{j,B}^\dag c_{j+r,A}^{}  \rangle= \begin{vmatrix}
        \zeta(r) & \zeta(r+1)  & \cdots & \zeta(r+n-1)\\
        \zeta(r-1) & \zeta(r)  & \cdots & \zeta(r+n-2)\\
        \vdots & \vdots & \ddots & \vdots\\
        \zeta(r-n+1) & \zeta(r-n+2)  & \cdots & \zeta(r)
    \end{vmatrix},
    \label{eqn:calculate_determinant}
\end{align}
where $\zeta(x)=\langle c_{j,B}^\dag
c_{j+x,A}^{}\rangle$ defined in Eq. (\ref{eqn:ciB(t)cjA(t)}).

\subsection{Generalization of the non-interacting SSH model result}
\label{app:generalised_ssh_op}
\begin{figure}[h]
    \centering
    \includegraphics[width=8cm]{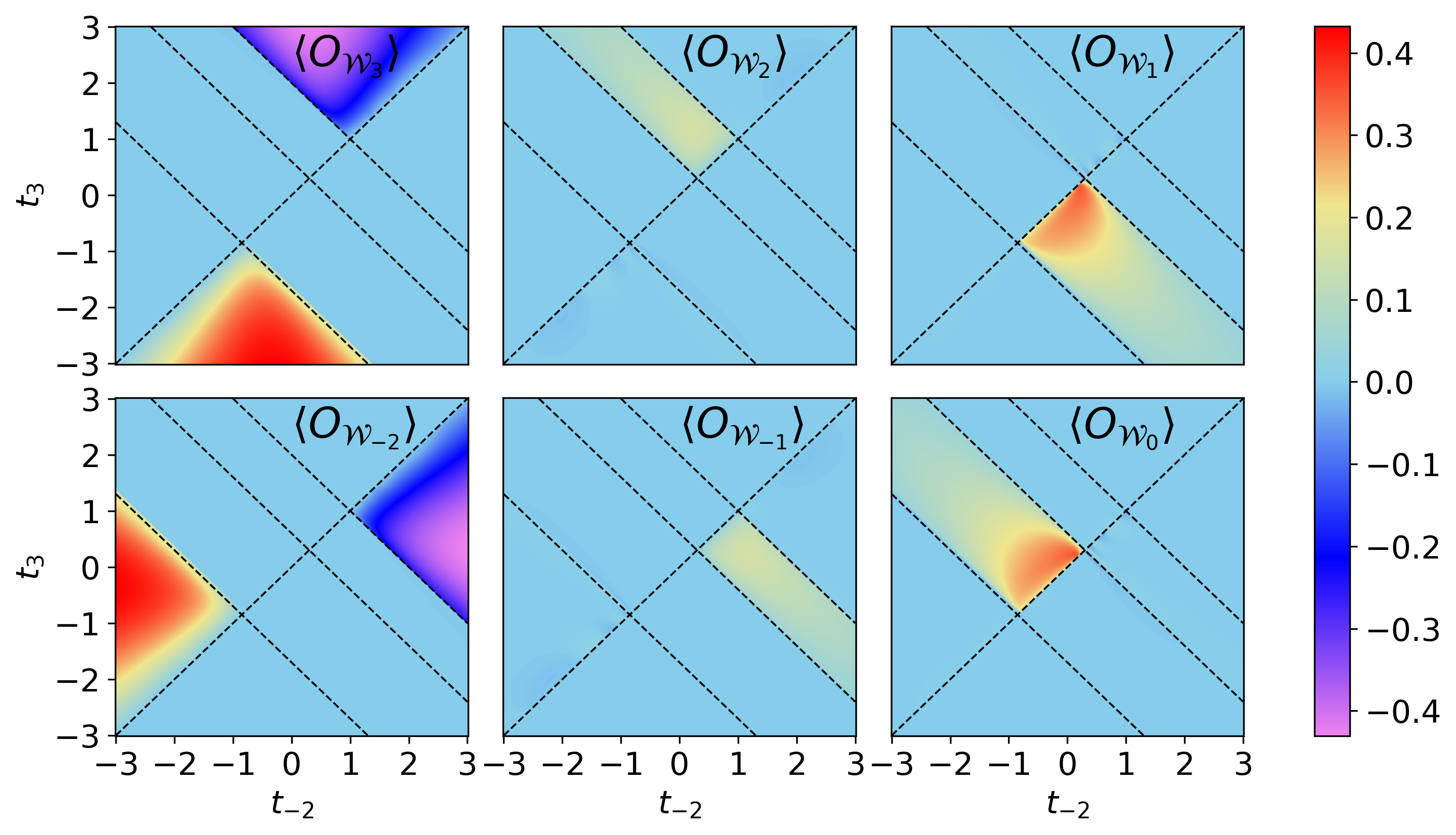}
    \caption{The OPs in Eq. (\ref{eqn:generalise_ssh_op}) as a function of $t_{-2}$ and $t_3$ with $t_0=t_1=-1$ and $t_{-1}=t_2=0$. The choice of $m$ for each OP is indicated in each panel. Black dashed lines indicate the phase boundaries determined by the winding numbers. Here, $N=80$ and $n=39$. The phase diagram can also be found in Ref. \cite{wong2024}. }
    \label{fig:generalise}
\end{figure}
We consider the generalized SSH model mentioned in the main text, namely, 
$$H = \sum_{j = 1}^{N} \sum_{i=\gamma}^{-\gamma}( t_i^{}c_{j,B}^\dagger c_{j+i,A}^{}  + \text{h.c.} )$$
For clarity, we identify the $t_a,t_b,t_c,t_d$ terms in Eq. (\ref{eqn:H_SSH}) as $t_0=t_a$, $t_1=t_b$, $t_2=t_d$, $t_{-1}=t_c$. 

To verify if the OP in Eq. (\ref{eqn:generalise_ssh_op}) is correct, we consider hoppings $\{t_{\ok{m}}\}_{\ok{m}\in\{-2,-1,0,1,2,3\}}$ to be non-zero while all other hoppings are set to zero. Figure \ref{fig:generalise} displays the expectation values of $O_{\mathcal{W}_{\ok  m}}$ in Eq. (\ref{eqn:generalise_ssh_op}).

\subsection{Domiant Fock states and the choice of OPs in the XXZ model}
\label{app:XXZ}
For the XY phase, there are a large number of Fock states with comparable weights. To gain intuition, we list some of the dominant ones ($\Delta=0$ and $L=16$):

\begin{gather}
\label{Fock:xy1}
\ket{1 0\ 1 0\ 1 0\ 1 0\ 1 0\ 1 0\ 1 0\ 1 0} \ \  \ |\xi_i|=0.062500, \\
\label{Fock:xy2}
\ket{\underline{0 0 \ 1 1}\  0 1\ 0 1\ 0 1\ 0 1\ 0 1\ 0 1} \ \ \ |\xi_i|=0.040045, \\
\label{Fock:xy3}
\ket{\underline{0 0 }\  1 0\ \underline{11}\ 0 1\ 0 1\ 0 1\ 0 1\ 0 1} \ \ \ |\xi_i|=0.034668, \\
\label{Fock:xy4}
\ket{\underline{0 0 }\ 1 0\ 1 0\  \underline{1 1} \ 0 1\ 0 1\ 0 1\ 0 1} \ \ \ |\xi_i|=0.032486, \\
\label{Fock:xy5}
\ket{\underline{0 0 }\ 1 0\ 1 0\ 1 0\ \underline{1 1} \ 0 1\ 0 1\ 0 1} \ \ \ |\xi_i|=0.031862, \\
\label{Fock:xy6}
\ket{\underline{0 0 \ 1 1\ 0 0\ 1 1}\ 0 1\ 0 1\ 0 1\ 0 1} \ \ \ |\xi_i|=0.027772, \\
\label{Fock:xy7}
\ket{\underline{0 0 \ 1 1}\ 0 1\ \underline{ 0 0\ 1 1}\ 0 1\ 0 1\ 0 1} \ \ \ |\xi_i|=0.026856, \\
    \label{Fock:xy8}
\ket{\underline{0 0 \ 1 1}\ 0 1\  0 1\ \underline{ 0 0\ 1 1}\ 0 1\ 0 1} \ \ \ |\xi_i|=0.026673, \\
\label{Fock:xy9}
\ket{\underline{0 0 }\  10\ \underline{11 \ 00 \ 11}\ 0 1\  0 1\ 0 1} \ \ \ |\xi_i|=0.025165, \\
\label{Fock:xy9.2}
\ket{\underline{1 1 \ 0 0 \ 1 1}\  0 1\ \underline{00}\ 1 0\ 1 0\ 1 0} \ \ \ |\xi_i|=0.025165, \\
\label{Fock:xy10}
\ket{\underline{0 0 \ 11 }\  010 \ \underline{11 \ 00}\ 10\ 10\ 1} \ \ \ |\xi_i|=0.024643, \\
\label{Fock:xy11}
\ket{\underline{0 0 \ 11\ 00 }\  10\  10 \ \underline{11}\ 10\ 10} \ \ \ |\xi_i|=0.024514, \\
\label{Fock:xy12}
\ket{\underline{0 0 \ 1 1}\ 0\ \underline{ 1 1 \ 0 0}\  1 0 \  1 0 \  1 0\ 1} \ \ \ |\xi_i|=0.024245, \\
\label{Fock:xy13}
\ket{\underline{1 1 \ 0 0}\ 1 0\ \underline{ 11}\   01 \ \underline{00}\  10 \  10} \ \ \ |\xi_i|=0.024170, \\
\label{Fock:xy14}
\ket{\underline{0 0 }\ 1 0\ \underline{ 11 \ 00}\   10 \ \underline{11}\ 01 \  01} \ \ \ |\xi_i|=0.023705, \\
\vdots \nonumber
\end{gather}


Each of the above represent several Fock states up to a global translation of one site. The largest weighted Fock state in the XY phase has the same configuration as that of the AFM phase [see Eqs. (\ref{Fock:xy1}) and (\ref{Fock:afm1010})]. However, the weight in the XY phase is relatively much smaller, i.e. the probability of finding such a state is only $0.0625^2\approx3.9\times10^{-3}$. For the other Fock states with smaller but comparable weights, their configurations contain some local violations to the AFM pattern in Eq. (\ref{Fock:xy1}) as underlined above.

\emph{$\langle O_{\rm{XY},2}\rangle$ is near zero in FM and AFM phase.---}Obviously, each term in Eq. (\ref{op:xy2}) will vanish in the expectation value of the FM GS (Eq. (\ref{Fock:fm})), since $\sigma_i^{\pm}\sigma_{i+j}^\mp\ket{0000000000000000}$ and $\sigma_i^{\pm}\sigma_{i+j}^\mp\ket{1111111111111111}$ are always zero for $j\ne 0$. For the AFM phase, although the superposition of Eq. (\ref{Fock:afm1010}) and Eq. (\ref{Fock:afm0101}) are not exactly the GS, those two states still dominate and we can neglect the contribution from the others. Therefore, the expectation value of each term in Eq. (\ref{op:xy2}) is almost vanishing in the AFM phase, since $\sigma_i^{\pm}\sigma_{i+j}^\mp\ket{1010101010101010}$ is neither equal to itself nor $\ket{0101010101010101}$.

\emph{$\langle O_{\rm{XY},2}\rangle$ being finite in XY the phase.---}In contrast, many dominant Fock states in the XY phase can be connected by the above definition of OP, hence giving rise to a non-zero $\langle O_{\rm{XY},2}\rangle$. 
For instance, $\sigma_2^+\sigma_3^-\ket{1010101010101010}=\ket{\underline{1100}101010101010}$ (see Eq. (\ref{Fock:xy2})). Another example is $\sigma_2^+\sigma_3^-\ket{\underline{00}10\underline{11}0101010101}=\ket{01\underline{0011}0101010101}$, which relates the Fock state in Eq. (\ref{Fock:xy3}) to that in Eq. (\ref{Fock:xy2}). Other than these nearest neighbor ``hoppings'', the next-nearest-neighbor or beyond are possible as well, as Eq. (\ref{op:xy2}) suggests. For example, $\sigma_{10}^+\sigma_8^-\ket{\underline{1100}10\underline{1100}101010}=\ket{\underline{1100}101\underline{0011}01010}$ (see Eq. (\ref{Fock:xy7}) and (\ref{Fock:xy10}) respectively). An example for ``hopping'' beyond the next-nearest-neighbor is $\sigma_{6}^+\sigma_9^- \ket{\underline{1100}1\underline{0011}0101010}=\ket{\underline{110011}01\underline{00}101010}$ (see Eq. (\ref{Fock:xy12}) and (\ref{Fock:xy9.2}) respectively). We remark that the longer the ``hopping'' is, the smaller the value for this particular choice, as less Fock states can be connected. 

\emph{$\langle O_{\rm{XY},4}\rangle$ being finite in the XY phase.---}To see that $\langle O_{\rm{XY},4}\rangle$ is also non-zero in the XY phase, let us first express the OP in terms of raising and lowering operators: 
\begin{align}
&\sigma^x_i\sigma^x_{j}\sigma^x_{k}\sigma^x_{l}\nonumber+\sigma^y_i\sigma^y_{j}\sigma^y_{k}\sigma^y_{l} 
    = \frac{1}{8}\Bigl( \sigma_i^+\sigma_j^+\sigma_k^+\sigma_l^+ + \sigma_i^-\sigma_j^-\sigma_k^-\sigma_l^- \nonumber\\&+ \sigma_i^+\sigma_j^+\sigma_k^-\sigma_l^- + \sigma_i^-\sigma_j^-\sigma_k^+\sigma_l^+  + \sigma_i^-\sigma_j^+\sigma_k^+\sigma_l^- + \sigma_i^+\sigma_j^-\sigma_k^-\sigma_l^+  + \sigma_i^-\sigma_j^+\sigma_k^-\sigma_l^+ + \sigma_i^+\sigma_j^-\sigma_k^+\sigma_l^- \Bigl).
\end{align}
The first two terms changes the total magnetization of the Fock states that are being acted on. However, since the ground state of the XY (and the AFM) phase is in the zero total magnetization subspace, the expectation values of these two terms must be zero. On the other hand, the $3^{\text{rd}}$ and the $4^{\text{th}}$, the $5^{\text{th}}$ and the $6^{\text{th}}$, as well as the $7^{\text{th}}$ and the $8^{\text{th}}$ terms give identical expectation value. Here are two examples illustrating the action of the $7^{\text{th}}$, and the $5^{\text{th}}$ term, respectively:
\begin{align*}
    \sigma_7^+\sigma_6^-\sigma_5^+\sigma_4^-|\underline{0011}010101010101\rangle&=|\underline{00}1010\underline{11}01010101\rangle,\\
    \sigma_6^+\sigma_7^-\sigma_5^+\sigma_4^-|0101\underline{0011}01010101\rangle&=|01\underline{0011}0101010101\rangle.
\end{align*}


\end{document}